\documentclass{moriond}

\def\Journal#1#2#3#4{{#1} {\bf #2}, #3 (#4)}

\def\JCP{\em J. Co. Ph. }
\def\MNRAS{\em MNRAS}
\def\apj{\em ApJ}

\def\be{\begin{equation}}
\def\ee{\end{equation}}
\def\bea{\begin{eqnarray}}
\def\eea{\end{eqnarray}}

\newcommand{\Photo}{}

\begin{document}
\vspace*{4cm}
\title{Formation and evolution of heavy sub-structures in the centre of galaxy clusters: the local effect of dark energy}

\author{ M. Arca Sedda$^1$, M. Donnari$^{1,2}$, M. Merafina$^1$ }

\address{\footnotesize{$^1$Department of Physics, Sapienza University of Rome\\
Piazzale A. Moro 5, I-00185 Rome, Italy\\
$^2$Department of Physics, University of Rome Tor Vergata\\
Via O. Raimondo , I-00173 Rome, Italy} }

\maketitle\abstracts{
We discuss how the centres of galaxy clusters evolve in time, showing the results of a series of direct N-body simulations. In particular, we followed the evolution of a galaxy cluster with a mass $M_{clus} \simeq 10^{14} $M$_{\odot}$ in different configurations. The dynamical evolution of the system leads in all the cases to the formation of dense and massive sub-structures in the cluster centre, that form in consequence of a series of collisions and merging among galaxies travelling in the cluster core. We investigate how the structural properties of the main merging product depends on the characteristics of those galaxies that contributed to its formation.}

\section{Numerical simulations}
We modelled the dynamical evolution of a galaxy cluster (GC) under the action of dark energy (DE) \cite{2016Donnari}, in order to understand whether its anti-gravitational term in the equation of motion of each particle may play a significant role in shaping the structure of the GC. We focused our investigation on the role of orbital decay and merging of galaxies moving in the cluster innermost region, aiming at finding whether DE can leave a footprint on the evolution of the cluster core.
We simulated GC composed of 240 galaxies using a modified version of a direct N-body code \texttt{HiGPUs} \cite{2013spera}. Each galaxy is modelled according to a Dehnen profile, whereas all the galaxies are distributed in the phase-space according to a King density profile. For DE we follow $\Lambda$CDM model and for the gaseous component (ICM) we use the so-called $\beta$-model.
In order to highlight the different role played by each component, we switched on/off
alternatively the DE or ICM, as indicated in Tab. \ref{tab:1}.
We found that galaxies moving within the Zero-Gravity Radius (ZGR), which is the radius within gravity overcomes antigravity, are mostly affected by mutual gravitational interactions, which drive the orbital decay of the most massive galaxies due to the action of dynamical friction (df) \cite{2014Mas}. This leads to an accumulation of heavy galaxies in the GC core, favouring their pairing and merging. In Fig. \ref{fig:fig1} (left panel) is shown how the df time-scale varies at varying galaxy mass and position.

\begin{table}
\caption{Left: Physical processes considered in simulations and their evolutionary ages. Right: Parameters of the simulated galaxy cluster.}
\label{tab:1}
\begin{minipage}{0.5\linewidth}

\vspace{0.4cm}
\begin{center}
\begin{tabular}{|c|c|c|c|}
\hline
Model & $\Lambda$ & ICM & $T_{ev} (Gyr)$ \\
\hline
MM1 & X & & 3.7 \\
MM2 & X & X & 5.3 \\
MM3 &  & X & 3.5 \\
MM4 &  & & 3.5 \\
\hline
\end{tabular}
\end{center}
\end{minipage}
\hfill
\begin{minipage}{0.5\linewidth}
\vspace{0.4cm}
\begin{center}
\begin{tabular}{|c|c|c|}
\hline
Parameter & Value \\
\hline
$M_{clus}$ & $9.2 \times 10^{13} M_{\odot}$ \\
$r_c$ & 0.1 M pc \\
$M_{g,min}$ & $9\times 10^{10} M_{\odot}$ \\
$M_{g,max}$ & $1.0 \times 10^{12} M_{\odot}$ \\
\hline
\end{tabular}
\end{center}
\end{minipage}
\end{table}

\section{Outcomes and conclusions}
Concerning the dynamical evolution of the cluster nucleus, we followed the orbits of the central galaxies. Indeed, these galaxies are efficiently affected by df, which drags them toward the cluster centre. Surprisingly, the combined action of
DE and ICM seems to facilitate collisions among galaxies. Indeed, the number of
merging is much higher in simulation MM2, in which our results show that a sequence of 21 merging and collisions occurs over a time-scale t $\simeq$ 5 Gyr driving the
formation of a huge spheroidal structure.
This massive central structure (MCS) has a dynamical mass of $6\times 10^{12} $M$_{\odot}$ and a central velocity dispersion of 1300 km/s.

\begin{figure}
\begin{minipage}{0.50\linewidth}
\centerline{\includegraphics[width=\linewidth]{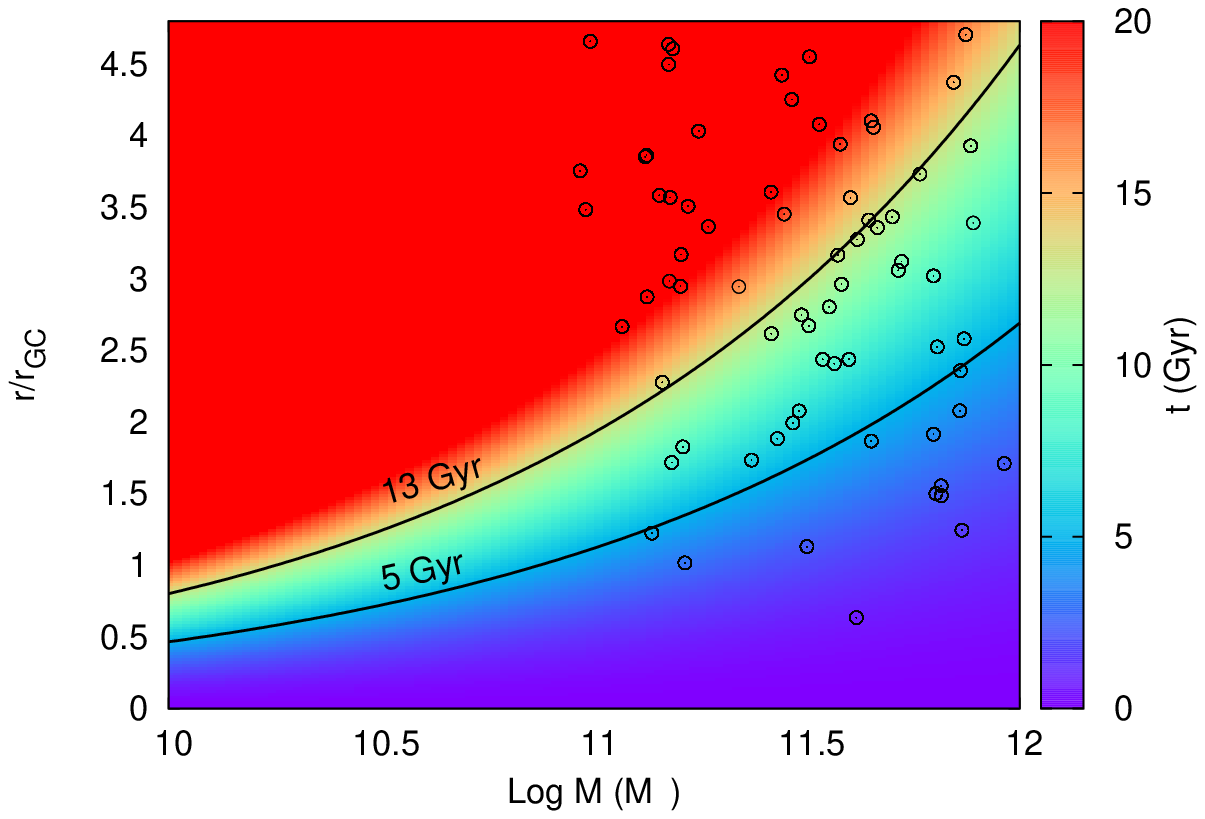}}
\end{minipage}
\hfill
\begin{minipage}{0.50\linewidth}
\centerline{\includegraphics[width=\linewidth]{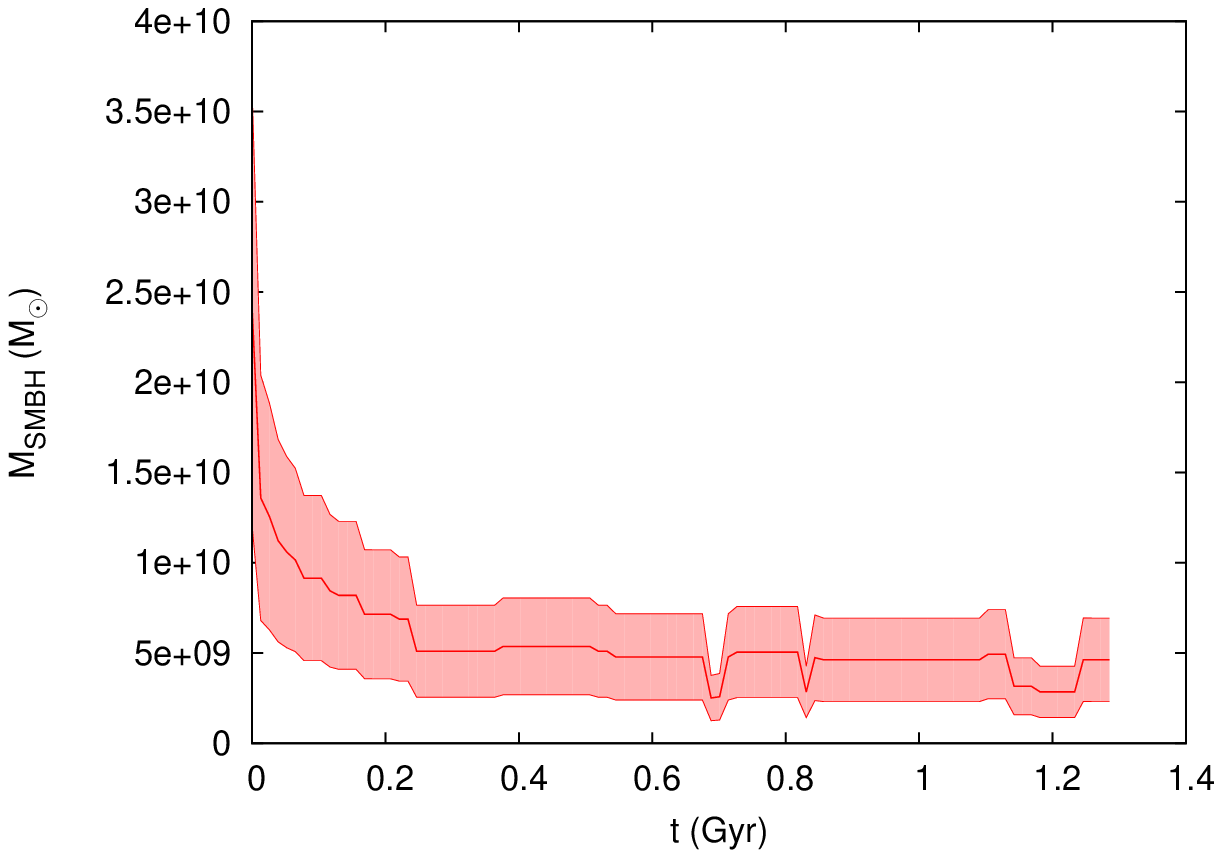}}
\end{minipage}
\caption{Left: Surface map of $\tau_{{\rm df}}$ at varying $M_g$ and $r$. The open circles represent the galaxies in our models. Right: Mass (with its error) of the SMBHs that move within $10$ kpc from the centre of the central structure.}
\label{fig:fig1}
\end{figure}

Another interesting outcomes of these simulations is the possibility to develop a study of super-massive black holes (SMBHs) dynamics. As widely accepted, the presence of SMBHs are thought to occupy the centre of the majority of obseved galaxies.
However, it is not completely clear how they interact during multiple galaxy mergings. To investigate this point, we assumed that an SMBH forms and grows efficiently at the centre of each merging galaxy before the MCS assembly ($\simeq$ 3 Gyr), and we simulated their mutual interactions and subsequent evolution in model MM2. We found that 13 of the 21 SMBHs are ejected away through multiple scatterings, reaching velocities up to $10^3$ km/s \cite{2014Miki}.
On the other hand, 4 SMBH sink to the MCS centre, where they form a bound system. Their subsequent evolution can lead to the formation of a final SMBH with mass $5\times 10^9 $M$_{\odot}$ as shown in Fig. \ref{fig:fig1}.

%=================== REFERENCES ======================

\end{document}